# Multimodal wearable EEG, EMG and accelerometry measurements improve the accuracy of tonic-clonic seizure detection in-hospital

Jingwei Zhang[1*], Lauren Swinnen[2*], Christos Chatzichristos[1], Victoria Broux[3], Renee Proost[4], Katrien Jansen[4,5], Benno Mahler[6], Nicolas Zabler[7], Nino Epitashvilli[7], Matthias Dümpelmann[7], Andreas Schulze-Bonhage[7], Elisabeth Schriewer[8], Ummahan Ermis[8], Stefan Wolking[8], Florian Linke[8], Yvonne Weber[8], Mkael Symmonds[9], Arjune Sen[9], Andrea Biondi[10], Mark P. Richardson[10], Abuhaiba Sulaiman I[11], Ana Isabel Silva[11], Francisco Sales[11], Gergely Vértes[12], Wim Van Paesschen[2,3**], Maarten De Vos[1,5**]

1 Department of Electrical Engineering (ESAT), STADIUS Center for Dynamical Systems, Signal Processing and Data Analytics, KU Leuven, Leuven, Belgium

2 Laboratory for Epilepsy Research, KU Leuven, Leuven, Belgium

3 Reference Center for Refractory epilepsy, Department of Neurology, UZ Leuven, Leuven, Belgium

4 Department of Pediatric Neurology, UZ Leuven, Leuven, Belgium

5 Department of Development and Regeneration, KU Leuven, Leuven, Belgium

6 Department of Neurology, Karolinska University Hospital, Stockholm, Sweden

7 Epilepsy Center, Department of Neurosurgery, Medical Center - University of Freiburg, Freiburg im Breisgau, Germany

8 Department of Epileptology and Neurology, University of Aachen, Aachen, Germany

9 Oxford Epilepsy Research Group, NIHR Biomedical Research Centre, Nuffield Department of Clinical Neurosciences, University of Oxford, Oxford, UK

10 Department of Basic and Clinical Neuroscience, Institute of Psychiatry Psychology and Neuroscience, King's College London, London, UK

11 Neurology Department, Coimbra Hospital and University Centre, Coimbra, Portugal

12 Epilepsy Seizure Detection – Neurology UCB Pharma, Brussels , Belgium

*, **: both authors contributed equally

## Correspondence

Jingwei Zhang, STADIUS Center for Dynamical Systems, Signal Processing and Data Analytics, Department of Electrical Engineering (ESAT), KU Leuven.

Address: Kasteelpark Arenberg 10, 3001 Leuven, Belgium.

Telephone number: 0032 16194874

E-mail address: jingwei.zhang@esat.kuleuven.be

**ORCID**

Jingwei Zhang: 0000-0001-8912-7339

Lauren Swinnen: 0000-0003-0531-9101

Number of words: 299 (<300) (abstract) + 3956 (<4000) (without acknowledgements, coi and ethical statement)

Number of references: 46 (<50)

Number of figures: 3

Number of tables: 3

## Summary

**Objective:** Most current wearable tonic-clonic seizure (TCS) detection systems are based on extra-cerebral signals, such as electromyography (EMG) or accelerometry (ACC). Although many of these devices show good sensitivity in seizure detection, their false positive rates (FPR) are still relatively high. Wearable EEG may improve performance; however, studies investigating this remain scarce. This paper aims 1) to investigate the possibility of detecting TCSs with a behind-the-ear, two-channel wearable EEG, and 2) to evaluate the added value of wearable EEG to other non-EEG modalities in multimodal TCS detection.

**Method:** We included 27 participants with a total of 44 TCSs from the European multicenter study SeizeIT2. The multimodal wearable detection system Sensor Dot (Byteflies) was used to measure two-channel, behind-the-ear EEG, EMG, electrocardiography (ECG), ACC and gyroscope (GYR). First, we evaluated automatic unimodal detection of TCSs, using performance metrics such as sensitivity, precision, FPR and F1-score. Secondly, we fused the different modalities and again assessed performance. Algorithm-labeled segments were then provided to a neurologist and a wearable data expert, who reviewed and annotated the true positive TCSs, and discarded false positives (FPs).

**Results:** Wearable EEG outperformed the other modalities in unimodal TCS detection by achieving a sensitivity of 100.0% and a FPR of 10.3/24h (compared to 97.7% sensitivity and 30.9/24h FPR for EMG; 95.5% sensitivity and 13.9 FPR for ACC). The combination of wearable EEG and EMG achieved overall the most clinically useful performance in offline TCS detection with a sensitivity of 97.7%, a FPR of 0.4/24 h, a precision of 43.0%, and a F1-score of 59.7%. Subsequent visual review of the automated detections resulted in maximal sensitivity and zero FPs.

**Significance:** In TCS detection with a wearable device, the main advantage of combining EEG with ACC, EMG or both was a marked reduction in FPR, while retaining a high sensitivity.

## Key Points

- TCS could be detected with high fidelity using wearable EEG recorded in a non-standard montage, e.g. behind-the-ear-EEG.

- Adding BTE-EEG to EMG, ACC or both improved the TCS detection performance compared to a single modality, by reducing FPR, while keeping sensitivity high.

- Consequent visual review of algorithm-labeled recordings by experts resulted in maximal sensitivity and no FPs.

- Thanks to its high computational efficiency, the current offline method can be a good basis for the development of an online real time detection system.

# 1 INTRODUCTION

Tonic-clonic seizures (TCS) are the most severe seizure type and can be of either focal or generalized onset. TCSs consist of two phases: during the tonic phase, there is stiffening of the muscles in the body, while in the clonic phase, jerking movements of the limbs are present. After the seizure has ended, the postictal period is characterized by a slow recovery from a comatose state, typically over 15-30 minutes. TCSs are the most important risk factor for sudden unexpected death in epilepsy (SUDEP) (1). It has been estimated that 69% of SUDEP cases in people who have TCS and live alone could be prevented if those individuals were attended at night or were free from TCS (2). Therefore, long-term home monitoring could be useful to improve detection and alert a caregiver to check on the patient whenever they have a TCS, consequently reducing morbidity and mortality (3, 4).

Although video-EEG is the gold standard for seizure detection, it is not suitable for long-term monitoring at home. To enable such long-term monitoring, a large variety of wearable TCS detection systems have been proposed (5, 6). Most wearable TCS detection systems are based on measurement of extra-cerebral signals, e.g., electromyography (EMG), accelerometer (ACC), electrodermal activity (EDA) and, electrocardiography (ECG).

EMG-based approaches have achieved high sensitivity, however, false positives (FPs) caused by muscle artifacts are relatively high, affecting their application in long-term monitoring (7-10). ACC, which reflects the changes in movement frequency and amplitude caused by TCSs, has also been widely used in TCS detection (11-14). The advantage of an ACC device is that it is small and requires no electrodes, making it more comfortable (13). In contrast, it is even more challenging for ACC-based approaches to avoid FPs, since certain daily activities often mimic the movement frequency and amplitude of seizures to produce seizure-like patterns e.g. while brushing teeth and playing cards (12). TCSs are also characterized by an increase in sympathetic nervous system activity, resulting in sweating and ictal tachycardia. EDA measures the electrical conductance of the skin, or sweating, and therefore, could be used for TCS detection (15, 16). However, EDA is limited by variable accuracy. Furthermore, ECG-based algorithms which could detect ictal tachycardia, have to be patient-specific due to the large variance in individual heart rate (17, 18).

A recent systematic review showed that ACC- and EMG-based systems have a high sensitivity but relatively high FPR, and suggested that future research focus on reducing FPR (19). To this end, EEG might perform better in TCS detection as it can directly monitor brain activity rather than indirect extra-cerebral biosignals. However, wearable EEG-based seizure detection remains challenging as it has to be unobtrusive, easy-to-wear, and not stigmatizing (20, 21). Moreover, such technology also suffers from muscle and movement artifacts when deployed in daily life.

Recently, our research showed the possibility of detecting seizures with only two discreet behind-the-ear (BTE)-EEG channels (22-25). The Sensor Dot (SD; Byteflies) is an unobtrusive multimodal wearable system measuring BTE-EEG, EMG, ECG, and movement through ACC and gyroscope (GYR) sensors (26).

This paper first investigates unimodal BTE-EEG in offline TCS detection and compares it with results of EMG, ACC and ECG. Secondly, we establish a simple multimodal TCS detection framework by combining the most relevant modalities. Finally, we use this multimodal detection algorithm to review the SD recordings and perform visual review of the automated detections to retain true positives (TPs) and discard FPs, so that TCS can be reliably logged and counted. In the future, we also aim to develop a real time detection system based on this research, which will require further development.

## 2 MATERIALS AND METHODS

### 2.1 Data collection

The dataset was obtained in the EIT-Health sponsored project SeizeIT2 (27). We recruited participants from 1 March 2020 to 30 June 2022 at the University Hospital Leuven, Freiburg University Medical Center, University Hospital Aachen, Coimbra University Hospital, Karolinska University Hospital, King's College Hospital London, and Oxford University Hospital. Participants were admitted to the video-EEG unit for recording of their habitual epileptic seizures as part of a pre-surgical evaluation or to further assess their epilepsy. Prior to enrollment, participants gave informed consent to have recordings with the SD in addition to standard full scalp EEG recording. The study was approved by all relevant ethical committees.

2.1.1   Experimental set-up

All the participants underwent video-EEG monitoring in the epilepsy monitoring unit (EMU) using the 25 electrodes full scalp EEG system according to the International Federation of Clinical Neurophysiology guidelines (28).

During the video-EEG monitoring, BTE-EEG, EMG, ACC, and ECG were recorded simultaneously with two SDs (Figure 1).

The SD is a small device of 24.5 x 33.5 x 7.73mm, 6.3 grams, with a battery life of 24 h and memory storage of 4 Gb. It consists of four electrodes and an Inertial Measurement Unit (IMU) that embeds the ACC and GYR sensors. Impedance was checked at the start of recording and kept below 5 kΩ throughout the recording. Set-up of the EEG and ECG/EMG SDs is visualized in Figure 1.

Both SDs were synchronized to enable multimodal analysis.

After 24h of recording, the SDs were placed in the docking station which allowed for the upload of the signals to the cloud and recharging of the battery.

2.1.2 Ground truth annotation

Ground truth annotations were based on the clinical video-EEG recordings. Neurologists (N.E., K.J., B.M., M.R., F.S., A.S-B., A.S., M.S., W.V.P., Y.W., S.W., U.E.), who were blinded to the wearable data and the output of the seizure detection algorithms, annotated all TCSs. The seizure annotations included the start and stop of the seizures, the start of the tonic phase, the start of clonic phase, and the type of the seizures (i.e., focal or generalized onset).

## 2.2. Methods

Multimodal seizure detection systems were developed by fusing state-of-the-art unimodal approaches.

2.2.1 Preprocessing and feature extraction

A. EEG

EEG signals were first bandpass filtered by a Butterworth filter in 1- 25 Hz range. Then, the filtered EEG signals were segmented into 2s windows with 50% overlap. Time domain, frequency domain, and entropy features were extracted from all bandpass filtered channels of each window.

Thereafter, the raw EEG signals were high-pass filtered by a Butterworth filter with a cut-off frequency of 1 Hz. The mean and normalized power between the 40 and 80 Hz bands of the high-passed EEG signals were calculated to act as the high-frequency features.

Core features of the EEG-based seizure detection have been detailed previously (23).

B. EMG

A new EMG algorithm was developed by selecting the features found in literature (7, 29, 30). The EMG signal was filtered by a high-pass Butterworth filter with a cut-off frequency of 20 Hz to remove the baseline drift and movement artifacts. The EMG signal was also segmented into 50% overlapping 2s windows.

In each window, 19 features found in literature were extracted. The importance of each feature was computed as the total reduction of the training criterion brought by that feature during the training of the random forest algorithm on an initial 10-participant dataset. The features were selected according to the computed feature importance. Five time-domain features and two entropy features were first designed to characterize the EMG activities. The time-domain features and entropy features were used to characterize the tonic and clonic phases.

C. ACC

The ACC recorded by the SD on the upper back of the participant was used for TCS detection. Three ACC channels measured the acceleration in the x axis, y axis and z axis. Each ACC channel was high-pass filtered with a Butterworth filter with the cut-off frequency at 2 Hz (due to the lower sampling frequency of SD). After that, a fourth channel was created by calculating the root mean square of those three channels. All the ACC signals were segmented into 2s windows with 50% overlap.

All the features used for ACC-based seizure detection were found in the literature (13). Besides the features from the time domain, features from Poincaré analysis (31) were also included to describe non-linear and dynamical information.

D. ECG

The ECG feature extraction method, as detailed in (32), first generates a R-peak series by detecting the R-peaks of ECG signal. With the R-peak series, six ECG features, i.e. the modified cardiac sympathetic index based on Lorenz, circadian rhythm-based feature, duration of the heart rate (HR) increase, HR before/after tachycardia, low-frequency power, and very-low frequency power, were then calculated.

Overall, all the unimodal systems were data-driven and feature-based. Table 1 lists the features selected for every modality.

2.2.2 Classification

A radial basis function (RBF) kernel support vector machine (SVM) was used to classify the seizure samples from non-seizure samples. SVM is a supervised learning method whose optimization objective is to find a hyperplane that maximizes the margin between the two classes. Considering the imbalanced nature of the seizure detection task, we under-sampled our non-seizure segments to train the SVM. The balanced training set included all the seizure samples and five times as many randomly sampled non-seizure samples.

Leave-one-patient-out nested cross-validation was applied to evaluate the seizure detection performance of every participant. In every fold, the two hyperparameters of SVM (i.e., the c and gamma of the RBF kernel) were tuned by an internal cross-validated grid search. The grid search generated candidate hyperparameter pairs from a parameter grid and evaluated their performance by conducting internal cross-validations on the training set. The hyperparameters that achieved the best performance were selected.

2.2.3 Modality fusion

To investigate the influence of multiple modalities, seizure detection was established with a multimodal late fusion strategy. Since TCSs tend to last around one minute (33), a detection was triggered when 20 consecutive windows (the windows last 2 seconds and have 1 second overlap) were classified as seizure in each modality. However, considering that artifacts may occur in some modalities, a threshold was allocated to tolerate some artifacts during the seizures. Namely, a detection will be triggered when a certain number of positive detections (36 positive detections for the combination of two modalities, 54 positive detections for the combination of three modalities) have been given in the 20 consecutive windows to tolerate artifacts. This strategy could decrease the FPR since it is not common for FPs to occur simultaneously.

2.2.4 Clinical validation

The original SD recordings, of which segments were flagged as TCS by the multimodal TCS detection algorithm, were visually evaluated by a neurologist (W.V.P.) and wearable data expert (L.S.). The algorithm-labeled multimodal files were presented in BrainRT (O.S.G.). The settings used for visualization of the signals were: a high-pass filter of 0,53 Hz, low-pass filter of 35 Hz and Notch filter of 50 Hz, and sensitivities of 100 µV/cm for EEG, 700 µV/cm for ECG and 100 µV/cm for EMG. The evaluators were blinded to the ground truth annotations and marked what they considered TPs and FPs. The sensitivity, precision and F1-score were calculated after the clinical review. Time needed to review the files was measured.

## 2.3 Evaluation

Event-based metrics were used to evaluate the seizure detection performance. Since the definition of TP, FP and false negatives (FN) has a great impact on the results, they are carefully defined for this real-life application.

- TP represents the correctly detected TCS; a seizure was correctly detected when the detection falls temporary within the ground-truth annotation. This means that a detection that starts before but ends after the start of a TCS was not considered a TP detection.

- FP represents the detections that happen outside of the duration of the TCSs.

- FN is an undetected TCS.

The sensitivity and FPR were calculated based on the above definitions of TP, FP and FN. Two additional metrics, i.e. precision and F1-score, were also calculated according to the definition in (23) to show the overall performance of the seizure detection methods.

## 3  RESULTS

## 3.1 Participant information

A total of 48 participants experienced TCS during study recording. However, the wearable recordings of 21 participants were incomplete, meaning that signals of one or more modalities were lost, either due to disconnecting of electrodes, transmission issues to the cloud when placing the SD in the docking station, or battery depletion of either SD. Since the different modalities could therefore not be fused, these recordings/participants were excluded. A total of 27 participants (16 females and 11 males) were included in this dataset. Twenty-five of them were adults, and two were pediatric. The average age of adult participants was 34 years (range: 20 - 54), while the average age of pediatric participants was 16 years (range: 15 - 17). A total of 3233 hours of recording was collected, and the average time of recording of each participant was 120 hours. The clinical characteristics of all participants included are listed in Table 2.

Forty-three focal to bilateral TCSs (FBTCS) and one generalized TCS (GTCS) were recorded during the study. The average duration of the seizures was 124 seconds (range: 55- 661 s).

## 3.2 Unimodal seizure detection

The SD multimodal measurements and different phases of detection of a TCS are shown in Figure 2. EEG, EMG, and ACC achieved a high sensitivity (Table 3). In contrast, the ECG only achieved a low sensitivity making it unsuitable for automated TCS detection.

EMG and ACC achieved high sensitivities but led to high FPRs. Since there were a high number of FPs for the EMG and ACC, their positive predictive value (PPV) (=precision) and F1-score were low. EEG achieved the highest sensitivity (100.0%) with a FPR (10.3/24hour) lower than EMG and ACC. This means that most TCSs were correctly detected with much fewer FPs, giving EEG the highest precision and F1-score among all unimodal modalities.

## 3.3 Multimodal seizure detection

The fusion strategy implies that the inclusion of a modality with low sensitivity in our multimodal framework will result in a lower sensitivity; hence, we opted to refrain from using ECG in any multimodal framework. The performance of all the combinations of EEG, EMG, and ACC was evaluated to investigate the added value of BTE-EEG in multimodal TC detection approach.

As shown in Table 3, the combination of modalities led to overall better performance than single modalities. All combinations led to a significant reduction in FPR compared with single modalities. The combination of non-EEG modalities (i.e. EMG and ACC) achieved 88.6% sensitivity and the FPR (0.8/24h). Compared with other modalities, BTE-EEG brought more substantial improvement; good performance was achieved in its combination with EMG or ACC. This is in line with the performance of the unimodal approaches: BTE-EEG was able to achieve high sensitivity with fewer FPs. The combination of EEG, EMG and ACC obtained the highest F1-score, combining a high sensitivity and PPV.

3.4 Clinical validation

We further presented the detections given by multimodal TCS detection algorithms to the two experts for a clinical review. On visual analysis, muscle activity was present on both EMG and EEG channels with some activation of ACC channels during the tonic phase (n=100%). The ensuing clonic phase was characterized by large amplitude changes of all ACC channels and high amplitude EMG activities in both EMG and EEG channels with a gradual decreasing frequency of muscle contractions (n=100%). In addition, we noticed ictal focal EEG changes and ictal tachycardia preceding the TCS in 68% and 48%, respectively. Overall, TCS and FPs were easily identified in the SD recordings (Figure 2). Sensitivity, therefore, remained maximal after the clinical review. The reviewers also found that multiple detections had arisen for the same event and therefore could easily filter them out. Both annotators successfully recognized all the FPs. The time needed to review the automatically detected TCS visually was on average **3.55 min** for a 24h file. In offline TCS detection, high sensitivity was preferred over low FPR since the FPs could be filtered out during the clinical review. Hence, we used the combination of EEG and EMG, which achieved, relatively, the best sensitivity of 97.7% and lowest FPR.

## 3.5. Computational efficiency

Both the EEG and EMG algorithms showed high computational efficiency in our study. Their average process time for a 2-seconds data window was 0.10 and 0.12 seconds, respectively, which is much shorter than the step size (1 second) between two data windows.

# 4  DISCUSSION

Several wearables to detect TCS based on EMG- or ACC measurement are currently available. The sensitivity of these devices is high, but FPR is also relatively high (19). Our main finding is that multimodal measurement of EEG, EMG and ACC can reduce FPR considerably. We have shown that each of these three modalities is sensitive in detecting TCS. By combining EEG with EMG or ACC, the sensitivity remained high, since TCS involve very typical patterns in these three signals. Automated detections in EEG, EMG or ACC, which were not confirmed in the other two signals, tended to be FP and were discarded. Previous theoretical research has shown that combining different modalities can outperform the unimodal counterparts in specific situations (34), which we confirm here. The multimodal detector using EEG, EMG and ACC had the highest F1-score (82.5% [0%: low sensitivity and PPV; 100%: optimal sensitivity and PPV]) followed by EEG and EMG (59.7%).

Our study focused on offline multimodal wearable detection of TCSs. In contrast to currently available wearables to detect TCS, our methodology allowed a visual review of recorded data. Visual review of automated detections enabled us to discard all FPs while maintaining maximal sensitivity. For visual review by experts, we used the EMG and EEG detector, since it had the highest sensitivity among the two detectors with the highest F1-score. Additionally, time needed to visually review the algorithm labels was 3.55 minutes for a 24-hour recording. This method could be used to log TCS in a reliable way, which could be used in clinical practice or anti-seizure medication trials.

Previously, Szabo and colleagues reported the best results using a unimodal EMG-based TCS detector (95.5% sensitivity, 0.02 FPR/24 hours) (35) and Kusmakar and colleagues with a unimodal ACC-based detector (95.2% sensitivity, 0.72 FPR/24 hours) (13). Our unimodal sensitivities for BTE-EEG, EMG, and ACC were comparable. The FPR of our unimodal EMG (31.9/24 hours), ACC (13.9/24 hours), and even EEG-based detectors (10.3/24 hours) were, though, strikingly higher. Even by reproducing the algorithm proposed in (36) on our dataset, the FPR remained high. This indicated poorer signal quality of our dataset (Figure 3). We observed that the baseline in the SD measurement is unstable and can make sudden jumps, which frequently coincided with FP detections. It would be of interest to study how much further FPR could be reduced when the signal quality of the multimodal detectors is optimized. It is also worth noting that the FNs were usually due to poor signal quality. For example, the EEG signals of patient 24 were partly corrupted into flat lines due to technical problems. The performance can be further improved if these technical problems were solved in the future.

The F1-score of the multimodal detector combining EMG of the deltoid muscle and ACC had a rather low F1-score of 41.5% and underperformed the multimodal combination including EEG. This could be because their FPs usually occur together as both modalities are related to body movement and signal quality is possibly poorer.

Visual analysis of our data, as in Figure 2, showed that EEG detection of TCS occurred during the tonic and clonic phase. In these phases, the EEG cerebral signal is largely obscured by prominent muscle artifacts. It is our interpretation that the unimodal EEG algorithm was mainly triggered by the scalp muscle EMG artifacts during the tonic and clonic phase. It highlights that the typical movement, which is captured from the scalp (through muscle activity and possible artifacts), is crucial for the accuracy of the TCS detector, and that scalp electrodes are better than deltoid electrodes for picking up such activity. Moreover, the actual cerebral activity on the EEG channels was used in the algorithm development to filter background activity and during visual review by experts as an additional characteristic of an FBTC to support their opinion.

## 4.1 Limitations and future work

Though our work shows the benefit of a multimodal detection framework consisting of EEG and EMG/ACC in TCS detection, there are important limitations.

First, this research was a phase 2 validation and thus consisted of collection of data within the hospital environment. In the future, we will further evaluate the proposed method outside of the hospital environment, which will be affected by more artifacts, potentially lowering the performance of the device (37).

Secondly, the detection algorithm could be further improved into a sequence-to-sequence model that could learn/recognize the typical evolution of TCSs with continuous muscle contraction during the tonic phase followed by a sequence of muscle contractions and relaxations with gradually decreasing frequency during the clonic phase.

Thirdly, the visual review time by experts of the SD recordings **was around 2 min**. To achieve a similar, short review time, training is needed to recognize the typical patterns in the different modalities.

Finally, the population was reduced due to the loss of at least one modality in 21 participants. Our future work will focus on developing a multimodal co-learning algorithm that is robust for missing modalities inputs. With a co-learning algorithm, the multimodal information in the training process could help improve the performance of the unimodal task at test time (38). Furthermore, some of the modalities were lost because of depletion of the battery of the SD. This could be avoided by either a notification from the device to the patient or by improving battery capacity. Finally, electrodes were sometimes disconnected due to movement during TCS. Further development of improved adhesive electrodes is therefore necessary.

Future research will focus on online detection of TCSSs. Previous work has investigated patients' requirements for an online seizure detection system. People with epilepsy desire a FPR between 1 to 2 per month if they have a low seizure frequency and a FPR between 1 to 2 per week if they have a higher seizure frequency (39). The combinations of BTE-EEG and EMG satisfies the desire of only having 1 to 2 FPs per week while keeping a high sensitivity. Our eventual aim is to develop a real-time detection system for TCS in order to alert caregivers. The high computational efficiency of the EEG and EMG methods indicates that the algorithm could run in real-time with the addition of a powerful processor or a real-time streaming to the cloud (40). For online detection, one could consider the combination of ACC, EMG and EEG, which had the best F1-score in our study, combining a high sensitivity with high PPV/precision. Unfortunately, the currently available SD EEG and EMG adhesive patches hamper the possibility of long-term monitoring over months due to limited signal quality and skin irritation. Besides providing long-term objective documentation of TCS frequency, wearables can also be used to assess disease severity (41, 42). TCS is a main risk factor for SUDEP, and an accurate TCS detector with online alarm function, can alert family members and caregivers. Various biomarkers known to be associated with a higher risk of SUDEP can be derived from the multimodal SD wearable detection device. Going forwards, it should be possible, for example, to study postictal generalized suppression (PGES) on the BTE-EEG, postictal tachycardia on the ECG, and respiration. If these data could be collected over longer periods of time in large cohorts, new insights into the pathophysiology of SUDEP could be obtained, and individual risk profiles could be developed based on these biomarkers (43-46).

In conclusion, we showed that unimodal detection with wearable EEG (BTE-EEG) outperformed the other modalities in TCS detection. Our results confirm that multimodal approaches, in particular those combinations with BTE-EEG, are able to improve the TCS detection performance over unimodal approaches by reducing FPR. This suggests that wearable BTE-EEG has the potential to play an important role in TCS detection in future research.

## Acknowledgement

The authors would like to thank all the participants for being involved in this research and Byteflies for their technical support. This research was funded by the EIT Health Grant: 21263 SeizeIT2 (Discreet Personalized Epileptic Seizure Detection Device). SeizeIT started out as an imec.icon research project (2016-2019), and continued as an EIT Health funded initiative (2020-2021). Project partners are KU Leuven, UCB Pharma, Byteflies, Pilipili, Neuroventis, Karolinska Institute, Aachen UMC, University Medical Center Freiburg, Oxford University Hospitals NHS Foundation Trust, King's College London and CHUC Coimbra. This research was also supported by the Flemish Government: "Onderzoeksprogramma Artificiële Intelligentie (AI) Vlaanderen" program and the "Deep, personalized epileptic seizure detection" FWO Research Project (G0D8321N); the Bijzonder Onderzoeksfonds KU Leuven (BOF): "Prevalentie van epilepsie en slaapstoornissen in de ziekte van Alzheimer" (C24/18/097).

## Disclosure of Conflicts of Interest

None of the authors has any conflict of interest to disclose.

## Ethical Publication Statement

We confirm that we have read the Journal's position on issues involved in ethical publication and affirm that this report is consistent with those guidelines.

## Contributions:

This project was conceived and supervised by M.D.V. and W.V.P.; J.Z. and L.S. wrote the manuscript with the comments from all the co-authors. L.S., V.B., R.P., K.J., B.M., N.Z., N.E., M.D., A.S-B., E.S., U.E., S.W., F.L., Y.W., M.S., A.S., A.B., M.P.R., A.S.I., A.I.S., F.S., G.V., W.V. P. contributed to the data collection and data annotation. J.Z. and C.C. developed the seizure detection algorithm.

Monitoring of Patients with Focal

Epilepsy Using two EEG channels. American Epilepsy Society (ASE) Annual Meeting 2022 2022; Nashville, Tennessee, U.S.A.

**Figures**

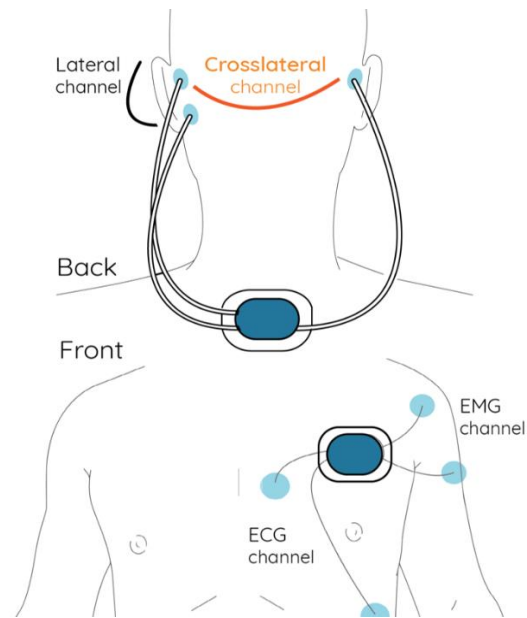

Figure 1. Set-up of Sensor Dot (SD). The SD system consists of multiple modalities using different 'dots' (recording device). To record EEG, a first dot is attached to the upper back of the participant and connected to the electrodes behind the ear. In cases of anticipated focal epilepsy (shown in figure), two electrodes are applied behind the ear on the side of the head with the epileptic focus and a unilateral channel is created. A third electrode is applied behind the other ear. By linking this with the upper electrode on the opposite side, a crosshead channel is created. In cases of anticipated generalized epilepsy, two electrodes are applied behind each ear and two unilateral channels are created. A second dot is attached to the chest of the participant and connected to the electrodes to record EMG on the deltoid muscle and ECG. All dots also inherently include an accelerometer and gyroscope to measure movement in different axes.

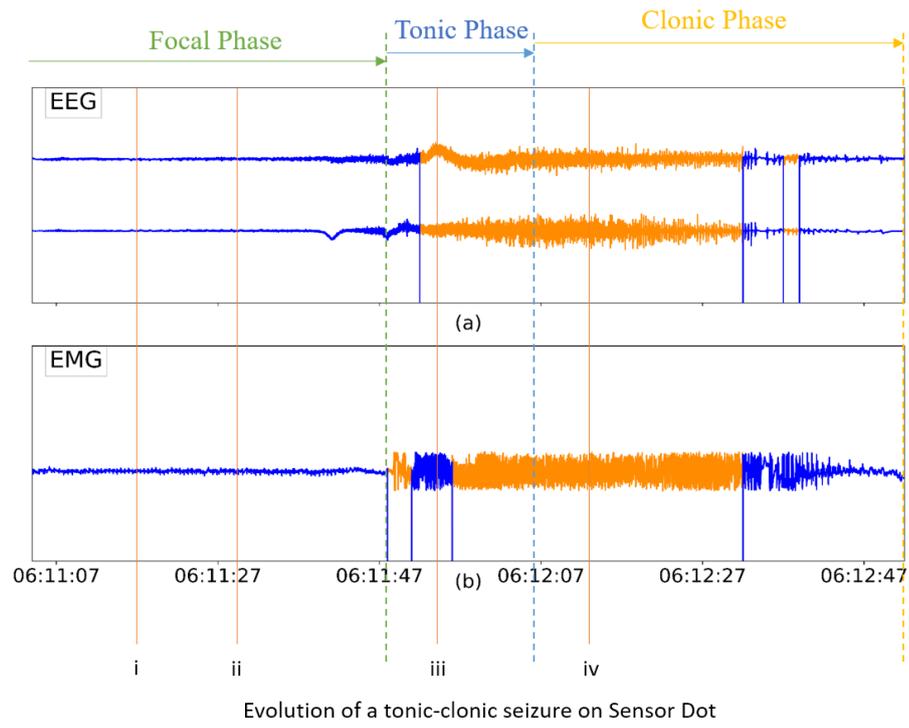

Figure 2. Demonstration of the multimodal detection on the TCS of patient 1. Top panel: (a) EEG signals during the TCS; (b) EMG signals during the TCS. Orange: the signal segments that had been detected as TCS by the algorithm. As more than 20 windows overlap, this is considered a TP detection. The different phases of the focal-to-bilateral tonic-clonic seizure (FBTCS) are defined on top of the figure with the focal phase, the tonic phase and the clonic phase. The signals that were not detected are in blue. Lower panel: the different stages of a TCS with settings for visual analysis: high-pass filter of 0,53 Hz, low-pass filter of 35 Hz and Notch filter of 50 Hz; and sensitivities of 100 µV/cm for EEG, 700 µV/cm for ECG and 300 µV/cm for EMG.(i) During seizure onset, tachycardia is visible on the ECG. (ii) During the focal part of the seizures, focal ictal discharges were visible on the EEG. (iii) During the tonic phase, muscle activity was present on EMG and EEG and there was no pronounced movement on the ACC. (iv) During the clonic phase, the jerking movements were visible on the ACC, as well as on EMG and EEG due to muscle artifacts, and decreased gradually in frequency.

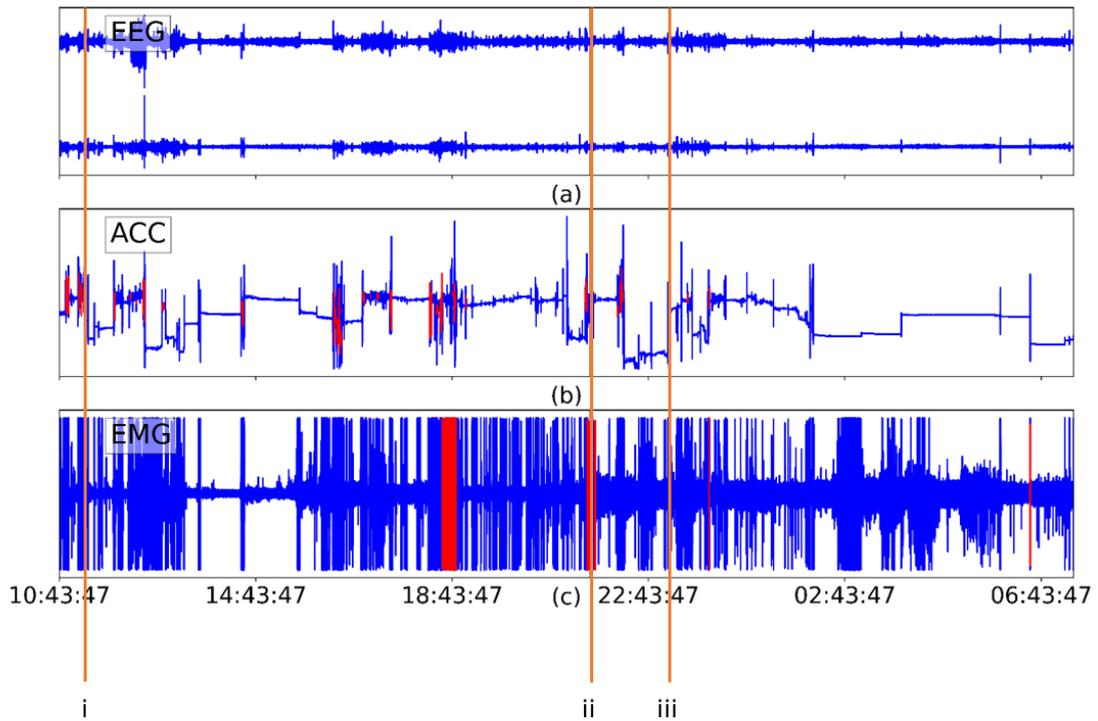

## Zoomed in examples of signals in different conditions

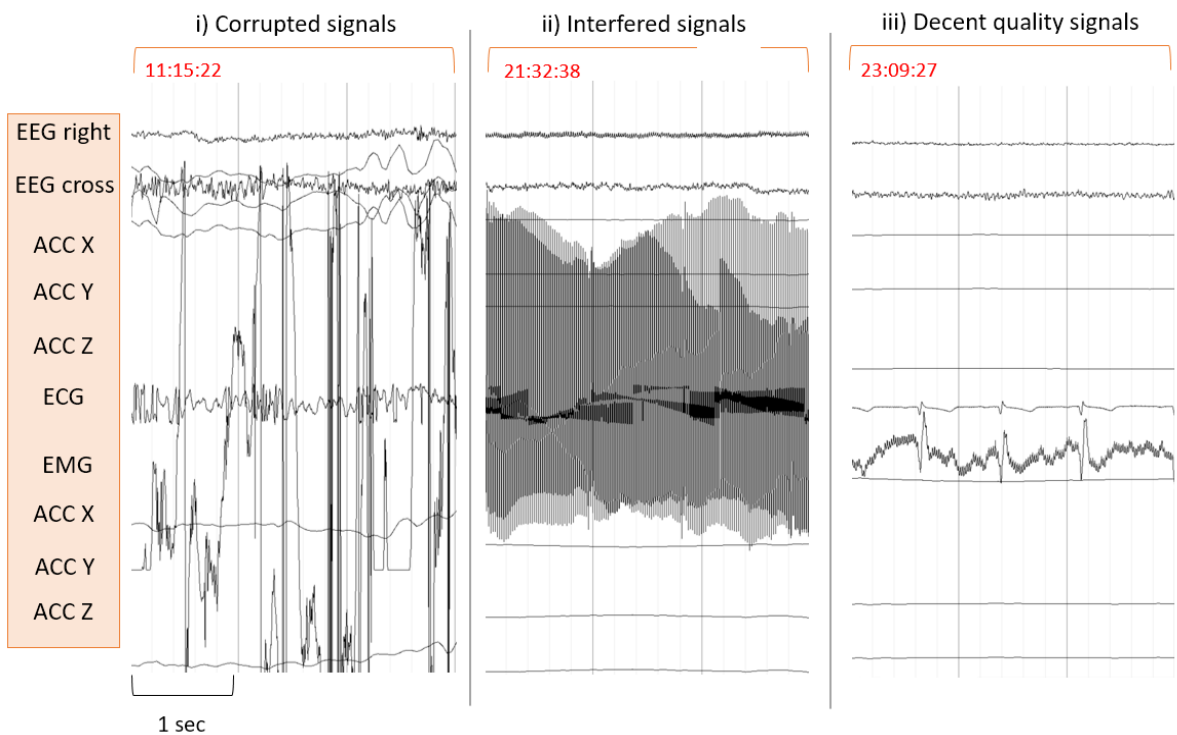

Figure 3. SD signal quality issues leading to FP detections. Example of multimodal signals in the first 24-h recording of patient 1. Top panel: (a): EEG, (b) ACC and (c) EMG signals of the entire recording. Red:

FP segments given by the algorithm. Lower panel: zoomed in examples of signals in different conditions: (i) corrupted EMG signals due to technical problems, (ii) interfered EMG signals due to high frequency interference, (iii) decent quality EMG signals. Notice that FPs usually occurred when signals were of poor quality ((i) and (ii)), especially when there was a high amplitude jump in the signals (i).

## Tables

Table 1. Features used in each modality

| EEG Features | |
| --- | --- |
| Time domain features | 1-3. Number of zero crossings, maxima, and minima; 4. Skewness; 5. Kurtosis; 6. Root mean square amplitude |
| Frequency domain features | 7. Total power; 8. Peak frequency; 9-16. Mean and normalized power in frequency bands: delta, theta, alpha, and beta; 17-18. Mean and normalized power in high-frequency band (40-80 Hz) |
| Entropies | 19. Sample entropy; 20. Shannon entropy; 21. Spectral entropy |

| EMG features | |
| --- | --- |
| Time domain features | 1. Zero crossing rate; 2. Wilson amplitude; 3. Mean absolute value; 4. Median absolute value; 5. Standard deviation |
| Entropies | 6. Cumulative residual entropy; 7. Fuzzy entropy |

| ACC features | |
| --- | --- |
| Time domain features | 1. Inter quartile range; 2. Standard deviation; 3. Kurtosis; 4. Skewness; 5. Zero crossing rate; 6-8. Maximum, Mean, Median amplitude |
| Frequency domain features | 9. Total average power |
| Poincaré analysis features | 10. SD1; 11. SD2; 12. Ratio; 13. CCM |

| ECG features | |
| --- | --- |
| Time domain features | 1. duration of the HR increase; 2. HR after tachycardia; 3.HR before tachycardia |
| Frequency domain features | 4. Very low frequency power; 5. Low frequency power. |

| ECG analysis features | 6. Modified cardiac sympathetic index based on Lorenz Plot; 7. Circadian rhythm-based feature |
|---|---|

Table 2. Participants information

| Participant | Age | Gender | Epilepsy syndrome | Etiology | Current ASMs | Previous ASMs | N of TCSs | Seizure Type |
|---|---|---|---|---|---|---|---|---|
| 1 | 25 | Female | Focal epilepsy (right; temporal) | Unknown | LEV | LCM, LEV, LTG, PER, TPM, VPA | 1 | FBTC |
| 2 | 28 | Male | Focal epilepsy (left; parietal) | Unknown | VPA, LTG, LCM | / | 1 | FBTC |
| 3 | 24 | Female | Focal epilepsy (right; unknown) | Unknown | BRV, LCM, | LEV, LTG, VPA | 1 | FBTC |
| 4 | 31 | Male | Focal epilepsy (left; frontal) | Structural (tumor) | PGB, BRV, CBZ, LEV | PER, TPM, VPA | 1 | FBTC |
| 5 | 38 | Male | Focal epilepsy (right; unknown) | Unknown | LCM, PER | BRV, LEV, LTG, PGB, TPM | 1 | FBTC |
| 6 | 34 | Male | Focal epilepsy (left; occipital/temporal) | Unknown | BRV | LEV, OCBZ | 1 | FBTC |
| 7 | 20 | Male | Other (both focal and generalized) | Unknown | LCM | LEV, LTG, VPA | 2 | FBTC |
| 8 | 38 | Female | Focal epilepsy (right; frontal) | Structural (other) | BRV, LCM | CLB, LEV, LTG, PGB, TPM, VPA | 1 | FBTC |
| 9 | 48 | Female | Focal epilepsy (left; temporal) | Unknown | / | CBZ, CLB, LEV, LTG, TPM, VPA | 1 | FBTC |
| 10 | 43 | Female | Focal epilepsy (left; temporal) | Unknown | CLB, LCM | LEV | 1 | FBTC |
| 11 | 35 | Male | Focal epilepsy (left; temporal) | Unknown | BRV | LEV, OCBZ, PER | 1 | FBTC |
| 12 | 39 | Female | Focal epilepsy (right; frontal) | Structural (tumor) | BRV, LCM | CLB, LEV, LTG, PGB, TPM, VPA | 1 | FBTC |
| 13 | 17 | Female | Focal epilepsy (left; frontal) | Unknown | BRV, CLB, LCM | LEV, OCBZ, TPM, VPA | 1 | FBTC |
| 14 | 15 | Male | Focal epilepsy (left; temporal) | Structural (cortical dysplasia) | LEV | CBZ, PER, VPA | 2 | FBTC |

| 15 | 30 | Female | Generalized epilepsy (GTCA) | Genetic | LCM, LTG, TPM | LCM, LEV, LTG, TPM | 1 | GTCS |
| 16 | 54 | Female | Focal epilepsy (left; temporal) | Structural (hippocampal sclerosis) | LTG | LEV, LTG, VPA | 2 | FBTC |
| 17 | 32 | Female | Focal epilepsy (left; central/complex) | Unknown | BRV, LCM, PHT | AZM, LEV, LTG, OCBZ, TPM, VPA | 1 | FBTC |
| 18 | 25 | Female | Focal epilepsy (left; frontal/temporal) | Structural (cortical dysplasia) | LEV, OCBZ, VPA | CBZ, LCM, LEV, OCBZ, PER, STM, TPM, VPA, ZNS | 1 | FBTC |
| 19 | 48 | Male | Focal epilepsy (right; temporal) | Structural (tumor) | BRV, CBZ | CBZ | 2 | FBTC |
| 20 | 42 | Male | Focal epilepsy (right; temporal) | Unknown | LEV, LTG, VPA, ZNS | CBZ | 2 | FBTC |
| 21 | 31 | Male | Focal epilepsy (left; temporal) | Structural (cortical dysplasia) | CLB, LTG, TPM, VPA | CBZ, LEV | 2 | FBTC |
| 22 | 21 | Male | Focal epilepsy (variable; temporal) | Infectious | OCBZ, TPM, VPA | CLB, CZP, LEV, LTG | 7 | FBTC |
| 23 | 38 | Female | Focal epilepsy (left; temporal) | Structural (vascular malformation) | CBZ, LTG | CBZ, LTG | 5 | FBTC |
| 24 | 26 | Female | Focal epilepsy (left; frontal/temporal) | Structural (vascular malformation) | LTG | CBZ, LEV | 2 | FBTC |
| 25 | 31 | Female | Focal epilepsy (right; temporal lobe) | Unknown | LCM, LEV | BRV, CBZ, LTG | 1 | FBTC |
| 26 | 31 | Female | Focal epilepsy (variable; temporal) | Unknown | CLB, LCM, LEV | LTG, OCBBZ, PER, VPA | 1 | FBTC |
| 27 | 33 | Female | Focal epilepsy (variable; temporal) | Unknown | OCBZ, PER, PGB | LCM, LEV, LTG, TPM, VPA | 1 | FBTC |

Abbreviations: AZM: acetazolamide; ASM: anti-seizure medication; BRV: brivaracetam; CBZ: carbamazepine; CLB: clobazam; CZP: clonazepam; FBTC: focal-to-bilateral tonic-clonic seizure; GTCS: generalized tonic-clonic seizure; LCM: lacosamide; LEV: levetiracetam; LTG: lamotrigine; OCBZ: oxcarbazepine; PER: perampanel; PGB: pregabaline; PHT: phenytoin; STM: sultiame; TPM: topiramate; VPA: valproate; ZNS: zonisamide.

Table 3. Results on seizure detection by the algorithm

|  | Sensitivity | FPR/24h | Precision | F1-score |
|---|---|---|---|---|
| ECG | 43.2% | 29.2 | 0.5% | 1.0% |
| ACC | 95.5% | 13.9 | 2.2% | 4.3% |
| EMG | 97.7% | 30.9 | 1.0% | 2.0% |
| EEG | 100.0% | 10.3 | 3.1% | 6.0% |
| ACC & EMG | 88.6% | 0.8 | 27.1% | 41.5% |
| EEG & ACC | 95.5% | 1.1 | 22.0% | 35.7% |
| EEG & EMG | 97.7% | 0.4 | 43.0% | 59.7% |
| EEG & EMG & ACC | 90.9% | 0.1 | 75.5% | 82.5% |